\documentclass[11pt,preprint]{aastex631}
\usepackage{amsmath}
\begin{document}
\title{Looking at infrared background radiation anisotropies with Spitzer: large scale anisotropies and their implications}

\author[0000-0003-2156-078X]{A. Kashlinsky}
\affiliation{Code 665, Observational Cosmology Lab, NASA Goddard Space Flight Center, 
Greenbelt, MD 20771, USA}
\affiliation{Dept of Astronomy, University of Maryland, College Park, MD 20742}
\affiliation{Center for Research and Exploration in Space Science and Technology, NASA/GSFC, Greenbelt, MD 20771, USA}
\email{Alexander.Kashlinsky@nasa.gov} 
\author[0000-0001-8403-8548]{Richard G. Arendt} 
\affiliation{Code 665, Observational Cosmology Lab, NASA Goddard Space Flight Center, 
Greenbelt, MD 20771, USA}
\affiliation{Center for Research and Exploration in Space Science and Technology, NASA/GSFC, Greenbelt, MD 20771, USA}
\affiliation{Center for Space Sciences and Technology, University of Maryland, Baltimore County, Baltimore, MD 21250, USA}
\author[0000-0002-3993-0745]{M.\ L.\ N.\ Ashby}
\affiliation{Center for Astrophysics $|$ Harvard \& Smithsonian, 60 Garden St., Cambridge MA 01238}
\author{J. Kruk}
\affiliation{Code 665, Observational Cosmology Lab, NASA Goddard Space Flight Center, 
Greenbelt, MD 20771, USA}
\author[0000-0002-8936-3582]{N. Odegard} 
\altaffiliation{deceased}
\affiliation{Code 665, Observational Cosmology Lab, NASA Goddard Space Flight Center, 
Greenbelt, MD 20771, USA}

\def\plotone#1{\centering \leavevmode
\epsfxsize=\columnwidth \epsfbox{#1}}

\def\wisk#1{\ifmmode{#1}\else{$#1$}\fi}

\def\wm2sr {Wm$^{-2}$sr$^{-1}$ }		
\def\nw2m4sr2 {nW$^2$m$^{-4}$sr$^{-2}$\ }		
\def\nwm2sr {nWm$^{-2}$sr$^{-1}$\ }		
\def\nw2m4sr {nW$^2$m$^{-4}$sr$^{-1}$\ }
\def\Ncut {$N_{\rm cut}$\ }
\def\lt     {\wisk{<}}
\def\gt     {\wisk{>}}
\def\le     {\wisk{_<\atop^=}}
\def\ge     {\wisk{_>\atop^=}}
\def\lsim   {\wisk{_<\atop^{\sim}}}
\def\gsim   {\wisk{_>\atop^{\sim}}}
\def\kms    {\wisk{{\rm ~km~s^{-1}}}}
\def\Lsun   {\wisk{{\rm L_\odot}}}
\def\Msun   {\wisk{{\rm M_\odot}}}
\def\um     { $\mu$m\ }
\def\sig    {\wisk{\sigma}}
\def\etal   {{\sl et~al.\ }}
\def\eg	    {{\it e.g.\ }}
\def\ie     {{\it i.e.\ }}
\def\bsl    {\wisk{\backslash}}
\def\by     {\wisk{\times}}
\def\cosec {\wisk{\rm cosec}}
\def\mic {\wisk{ \mu{\rm m }}}

\def\amin   {\wisk{^\prime\ }}
\def\asec   {\wisk{^{\prime\prime}\ }}
\def\cc     {\wisk{{\rm cm^{-3}\ }}}
\def\deg     {\wisk{^\circ}}
\def\ddeg   {\wisk{{\rlap.}^\circ}}
\def\damin  {\wisk{{\rlap.}^\prime}}
\def\dasec  {\wisk{{\rlap.}^{\prime\prime}}}
\def\approxeq{$\sim \over =$}
\def\abouteq{$\sim \over -$}
\def\percm{cm$^{-1}$}
\def\percmsq{cm$^{-2}$}
\def\percmcub{cm$^{-3}$}
\def\perhz{Hz$^{-1}$}
\def\perpc{$\rm pc^{-1}$}
\def\persec{s$^{-1}$}
\def\peryr{yr$^{-1}$}
\def\te{$\rm T_e$}
\def\tenup#1{10$^{#1}$}
\def\to{\wisk{\rightarrow}}
\def\thin{\thinspace}
\def\uk{$\rm \mu K$}
\def\p{\vskip 13pt}

\begin{abstract}
We use Spitzer/IRAC deep exposure data covering two significantly larger than before sky areas to construct maps suitable for evaluating source-subtracted fluctuations in the cosmic infrared background (CIB). The maps are constructed using the self-calibration methodology eliminating artifacts to sufficient accuracy and subset maps are selected in each area containing approximately uniform exposures. These maps are clipped and removed of known sources and then Fourier transformed to probe the CIB anisotropies to new larger scales. The power spectrum of the resultant CIB anisotropies is  measured from the data to $>1^\circ$ revealing the component well above that from remaining known galaxies on scales $>1\arcmin$. The fluctuations are demonstrated to be free of Galactic and Solar System foreground contributions out to the largest scales measured. We discuss the proposed theories for the origin of the excess CIB anisotropies in light of the new data. Out of these, the model where the CIB fluctuation excess originates from the granulation power due to LIGO-observed primordial black holes as dark matter appears most successful in accounting for all observations related to the measured CIB power amplitude and spatial structure, including the measured coherence between the CIB and unresolved cosmic X-ray background (CXB). Finally we point out the use of the  data to probe the CIB-CXB cross-power to new scales and higher accuracy. We also discuss the synergy of these data with future CIB programs at shorter near-IR wavelengths with deep wide surveys and sub-arcsecond angular resolution as provided by Euclid and Roman space missions.
\end{abstract}

\section{Introduction}
\label{sec:introduction}

The cosmic infrared background (CIB) contains important information on emissions and populations throughout the cosmic ages \citep{Bond:1986}. 
Because of the smoothness of the Solar System (zodiacal light) and to a smaller degree Galactic (ISM, cirrus) foregrounds, its anisotropies are more identifiable than the mean levels.  
In the near-IR, the CIB anisotropies may reflect emission from early sources inaccessible to direct telescopic studies \citep{Cooray:2004,Kashlinsky:2004}. The follow-up measurements using deep {\it Spitzer} IRAC data revealed significant source-subtracted anisotropies at 3.6 and 4.5 \mic, first found by \cite{Kashlinsky:2005a}. These exceed the anisotropies from known remaining galaxy populations \citep{Kashlinsky:2005a,Helgason:2012a}. This signal has been confirmed in further {\it Spitzer} measurements \citep{Kashlinsky:2007a,Kashlinsky:2012,Cooray:2012} and its origin is a subject of intense discussions \citep[see review by][]{Kashlinsky:2018}. The CIB fluctuations appear uncorrelated with the optical HST galaxies out to $m_{\rm AB}\gtrsim 28$ \citep{Kashlinsky:2007}, consistent with their origin at  high-$z$ or in more nearby (very) faint sources.
The {\it Spitzer}-found source-subtracted CIB fluctuations are in agreement with those subsequently probed by {\it Akari} \citep[][]{Matsumoto:2011,Seo:2015} \citep[see Fig. 20 in][]{Kashlinsky:2018}, which extended the measurement to 2.2 \mic\ and suggested the Rayleigh-Jeans spectral dependence for sources producing the fluctuations over the observer 2-5 $\mic$ range, $\nu I_\nu \propto \lambda^{-3}$.   Importantly, any successful interpretation must account for the observed high coherence between the near-IR source-subtracted CIB and the unresolved soft X-ray [0.5-2] keV background \citep[][]{Cappelluti:2013,Mitchell-Wynne:2016,Cappelluti:2017,Li:2018}, which indicates high proportions of accreting black holes (BHs) among its sources and cannot be accounted for by the populations found at low/intermediate $z$ \citep[][]{Helgason:2014}.

In this {\it Letter} we reduce the Spitzer/IRAC data from two large areas integrated sufficiently deeply to measure source-subtracted CIB fluctuations to angular scales not probed before. After evaluating and subtracting the instrument noise we find CIB anisotropies on degree scale which cannot be accounted for by remaining known galaxy populations and/or Solar System and Galactic foregrounds. On smaller scales the new measurements, achieving higher statistical accuracy than before, isolate the CIB components from remaining known galaxies, which will be a subject of forthcoming separate discussion.
We discuss implications of the measured CIB signal for the proposed models for the new populations. It appears that the currently best explanation of the signal and its properties lies with the LIGO-type primordial black holes (PBHs) making up the dark matter (DM) where the inevitable extra granulation component dominates the inflation-produced power in matter fluctuations at high $z$ and sub-galactic mass-scales, thus increasing early abundance of the necessary collapsed halos. We conclude with discussing the future important venues, achievable with the dataset we present here, which encompass new CIB-CXB probes, synergy with the current Euclid-based CIB science, and prospects with the forthcoming surveys by the NASA Roman mission to be launched in 2026.
\section{Spitzer Data and Processing}

For this study we used subsets from public data of the Spitzer IRAC 3.6 and 4.5 $\mu$m 
observations from Program IDs 13153 and 13058 \citep{Capak:2016}. 
One field, centered near the north ecliptic pole (NEP), corresponds to the Euclid Deep Field-North [NEP, Galactic coordinates $(l, b) \approx (95.8\arcdeg, 29.9\arcdeg)$]. The other field, centered near the Chandra Deep Field South (CDFS), matches the Euclid Deep Field-Fornax [CDFS, Galactic coordinates $(l, b) \approx (224.1\arcdeg,-54.6\arcdeg)$].
The data were self-calibrated using the method of \cite{Fixsen:2000}
as described by \cite{Arendt:2010}. 
The self-calibrated data were mapped into sky maps with a $1\farcs2$ pixel scale. In each of the two regions we then selected areas with approximately uniform exposure, $t_{\rm exp}$. For the NEP region, a square region of $4096 \times 4096$ pixels 
($4915\farcs2\times 4915\farcs2$) was selected from the center of the field for the power spectrum analysis. Hereafter, for historical reasons, the selected region is referred to as NEP. 
In the CDFS field, a rectangular region of $6200\times 2850$ pixels
($7440''\times 3420''$) was selected to avoid the portion of the field 
that was missing from these observations due to the prior  deeper Spitzer observations \citep{Dickinson:2003} where the CIB anisotropies have been probed at lower shot-noise levels but out to significantly smaller angular scales \citep{Kashlinsky:2007a}.
Hereafter, the selected region is referred to as CDFS in the figures and below. 
The range of $t_{\rm exp}$ in each of the selected regions is shown for each band in the Appendix.

As is now standard, we first mask sources iteratively at pixels with flux in pixel $i$ exceeding $F_i\geq \langle F\rangle + N_{\rm cut}\sigma_F$ with $N_{\rm cut}$=3 and remove additional $3\!\times\!3$ pixels around such pixel to account for the beam. The procedure converged after about 20--25 iterations. The common mask between the two wavelengths is then used in the final analysis. The fractional area left for the Fourier analysis is $f_{\rm sky}=0.61$ for the NEP field,  and $f_{\rm sky}=0.63$ for the CFDS. The Appendix 
shows the resultant maps.

In addition, an iterative procedure was applied to 
remove  the extended wings of both resolved and unresolved sources \citep{Arendt:2010}. 
For the present investigation, focused on probing the over-arcminute scales CIB anisotropies, we found that
subtraction of this contribution, at any number of iterations, had at most a negligible effect 
on the large scale background fluctuations. Hence, for brevity in the results here, the model was used to only, at most, a limited number of iterations with the goal of attaining the 
shot noise levels presented below. In terms of sources, in both final (clipped) maps the flux dispersion corresponds to the same $3\sigma$ magnitude $m_{\rm AB}=24.8,24.5$ at 3.6, 4.5 \mic\ respectively.
The dependence of the subarcminute CIB fluctuations arising from known galaxy populations, and specifically isolating the CIB fluctuation contributions from known populations as function of depth and the physics implied there, will be addressed in a forthcoming publication. 

The images, masked of the
clipped sources, yield the fluctuation field, $\delta
F(\vec{x})$ (see Appendix for distribution of $\delta
F$). Its Fourier transform (FT),
$\Delta(\vec{q})= \int \delta F(\vec{x})
\exp(-i\vec{x}\cdot \vec{q}) d^2x$, is
then calculated. The 1-D power spectrum on the angular scale $2\pi/q$ is $P(q)=\langle |\Delta(\vec{q})|^2\rangle/f_{\rm sky}$, with the azimuthal average taken over
all the Fourier elements at the given $q$. 
To enable easier comparison with previous results, the power spectrum was evaluated at spatial frequency ($q$) bins selected in \cite{Kashlinsky:2012}. The Appendix 
shows the number of independent Fourier pixels, $N_q$, at the selected values of $q$, and the standard deviation (dispersion) in the evaluated value of $P(q)$ is then $P(q)/\sqrt{N_q}$. 

As in previous studies, we construct the noise map by separating 
the total set of individual exposures into two subsets, producing 
separate images, and then subtracting the two images, thus removing stable celestial sources, but leaving the noise properties of the data. 
For these particular observations some extra steps were required as described in the Appendix leading to self-consistent and robust noise independent at each wavelength.

\section{CIB anisotropies}

\begin{figure}[ht!]
\includegraphics[width=6in]{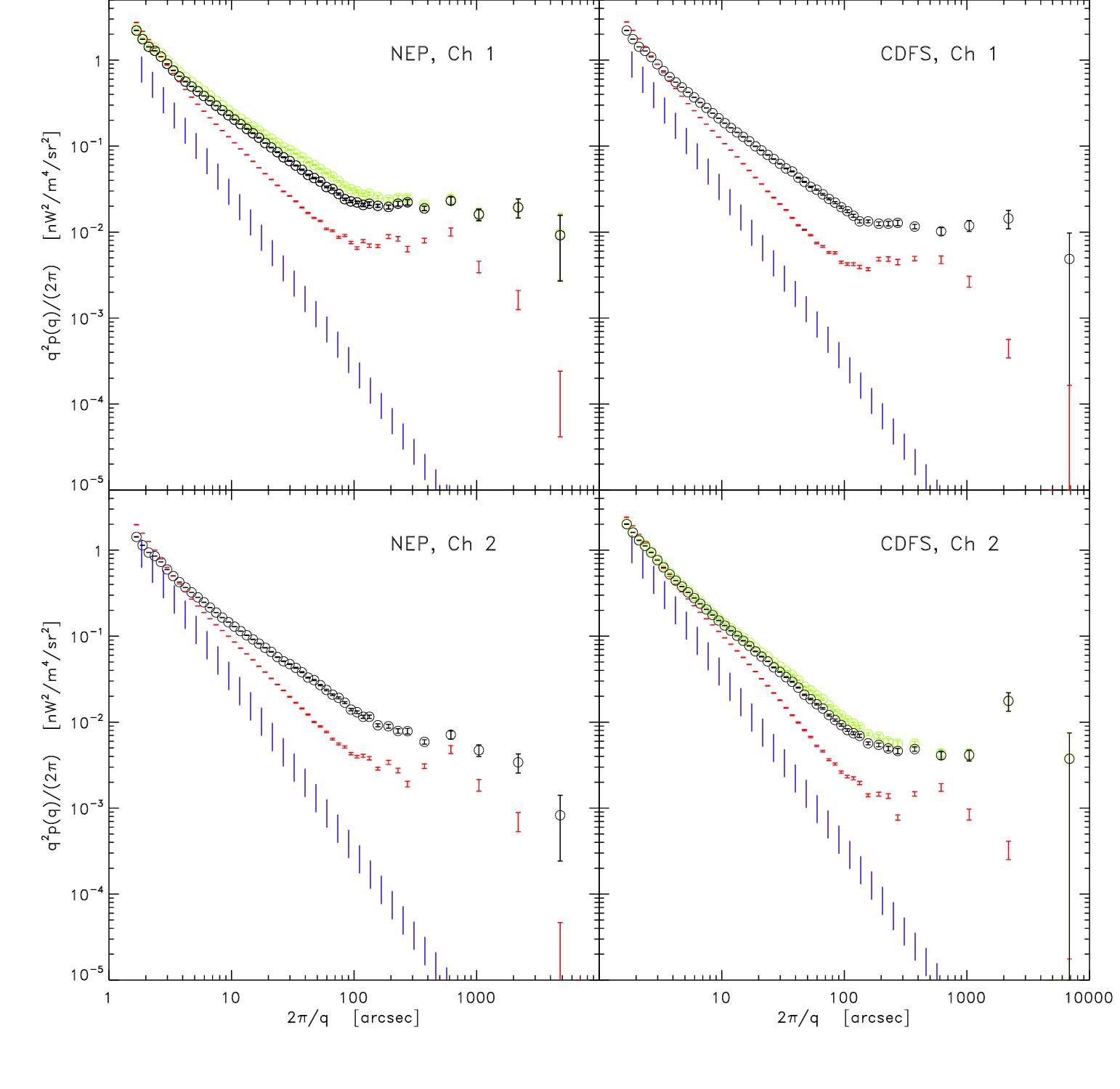}
 \caption{Open circles with error bars show the diffuse light fluctuations in the final maps. Black circles show the signal at the final model subtraction adopted here and the green circles show results without model subtraction when the subtraction was performed (NEP Ch1, CDFS Ch2); the subtraction was performed to achieve close levels of the effective shot noise between the 2 fields. The differences are negligibly small above arcminute scales which are the subject of the study here. Red error bars show the noise power estimates from the $(A-B)/2$ maps. The vertical blue dashes show the estimated range of the zodi contributions to the white noise per \cite{Arendt:2016}; the contribution is small at angular scales outside the beam.
  }
\label{fig:fig2}
\end{figure}

Figure \ref{fig:fig2} shows the resultant spatial fluctuations of the maps and the noise. The diffuse light fluctuations significantly exceed the noise at all scales outside the instrument beam with the excess coming from the cosmological background and/or Solar System and Galactic foregrounds.
Zodiacal light contributions can be deduced from \cite{Kelsall:1998} and \cite{Arendt:2016}. 
Following the methods used in \cite{Kashlinsky:2019}, 
the cirrus contribution of the ISM to the observed power spectrum 
was estimated using the 100\mic\ IRAS data \citep{Neugebauer:1984}. 

In a previous experiment, no correlation between the larger scale 
fluctuations and zodiacal light was detected \citep{Arendt:2016}.
However, there was a correlation found between the zodiacal light intensity and 
the white noise level (estimated in 4 different ways). 
If we apply those correlations to the present observations with the 
mean values of the ZODY\_EST keywords from each of the individual exposures, 
we find the zodiacal light contributing white noise levels shown for each field and wavelength as blue dashes in Fig. \ref{fig:fig2}.
The ranges in the estimates are from the different methods of calculating 
white noise levels. There is relatively little difference 
in the expected zodiacal light brightness between the two fields, because the 
zodiacal light variations are modest between
ecliptic latitudes $45\arcdeg$ and $90\arcdeg$ \citep{Levasseur-Regourd:1980,Leinert:1998,Kelsall:1998}. 

\begin{figure}[ht!]
\includegraphics[width=5.25in]{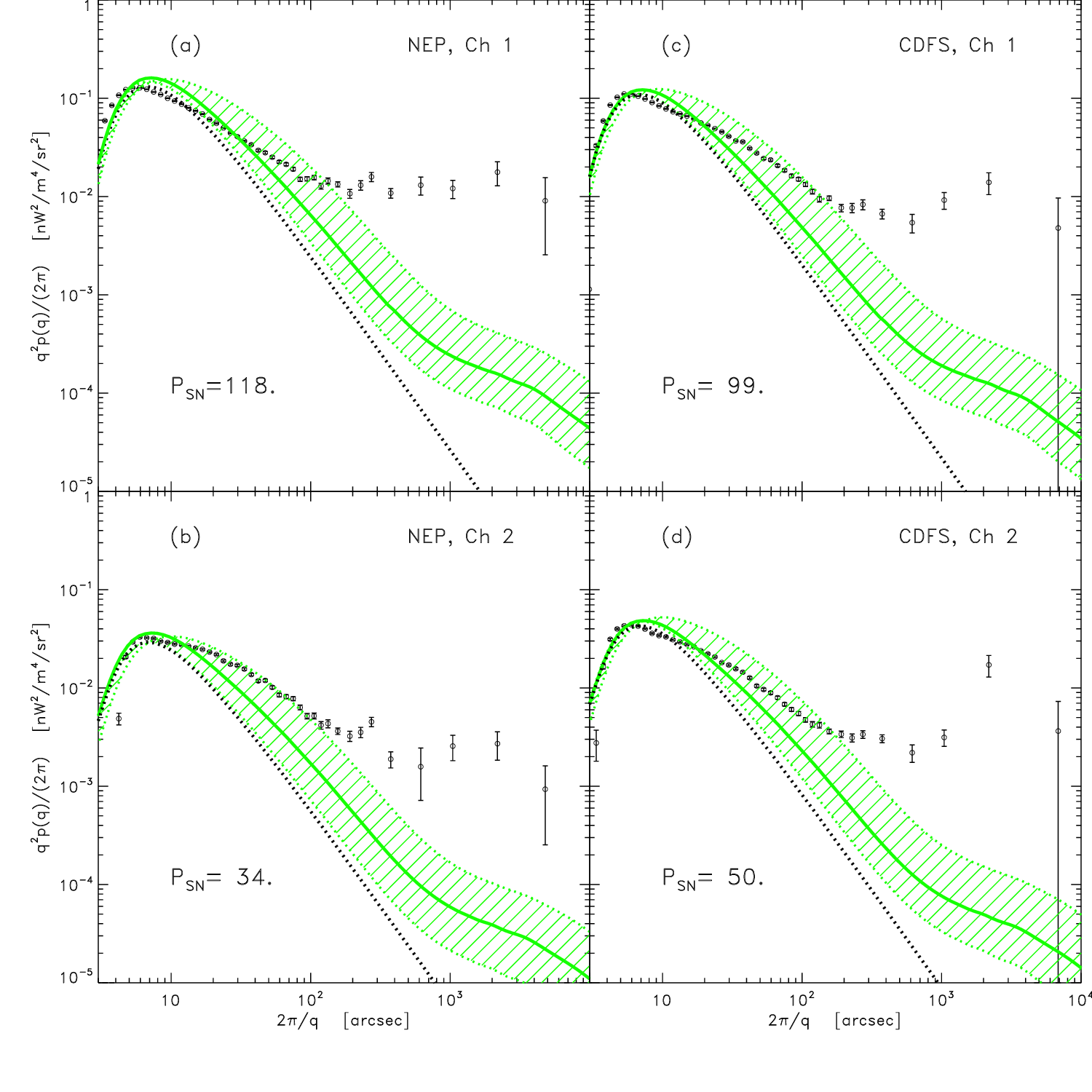}
 \caption{Source-subtracted CIB fluctuations after noise subtracted with errors propagated. Dotted line is the shot noise fitted to small scale data; $P_{\rm SN}$ is marked in nJy$\cdot$nW/m$^2$/sr. Green lines are the \cite{Helgason:2012a} reconstruction with shot-noise as fitted plus 1H and 2H terms, all convolved with the beam; solid is the default reconstruction and dotted lines are the allowed span.}
\label{fig:fig4}
\end{figure}
The ISM/cirrus estimates are made using 100 $\mu$m IRAS data \citep{Neugebauer:1984}, which fully cover both of 
the studied fields.
Both ISSA and IRIS \citep{Miville-Deschenes:2005} versions of the IRAS 
survey images yield similar power spectra. Given the uncertainty in the scaling factors from the IRAS to the IRAC wavelengths, we computed the cross-power, $P_{{\rm ISM}\times{\rm IR}}$ between the 100 \mic\ IRAS maps and the IRAC IR maps, both subject to the masking here, together with the ISM and IR autopowers, $P_{\rm ISM}$ and $P_{\rm IR}$. The quantity $P_{{\rm ISM}\times{\rm IR}}/\sqrt{P_{\rm ISM}P_{\rm IR}}$ gives an estimate of the amplitude of possible contributions of ISM to the IR/CIB power measured here.
The Appendix shows that this fractional cross power is small, and thus we find no evidence that
ISM emission contributes to the source-subtracted background power spectrum 
in these fields. 
Further indications of lack of ISM emission are that, as discussed below, there appears 
good agreement of the large scale fluctuation spectrum between fields at 
various Galactic latitudes, and that this large scale power is flatter than
would be expected from ISM emission, which is typically steeper with 
$P_{\rm ISM} \propto (2\pi/q)^\alpha$ where $\alpha \approx 2$-$3$
\citep[e.g.][]{Gautier:1992, Wright:1998, Miville-Deschenes:2002, Kiss:2003,
Lagache:2007, Bracco:2011, Penin:2012}.

The noise power was subtracted from the net map power to give the resultant cosmological CIB fluctuations. 
Figure~\ref{fig:fig4} plots the final source-subtracted CIB fluctuations (from the maps in the Appendix) at each region and wavelength. The results are mutually consistent between the 2 regions as required by their cosmological origin. (The excess power at $\sim 2000\arcsec$ in the CDFS fluctuations at 4.5 \mic\ is explained in the Appendix.)

\section{Cosmological implications}

The dotted lines in Fig.~\ref{fig:fig4}a--d show the {\it effective} shot-noise component, with the equivalent power marked in each panel in nJy$\cdot {\rm nW}\,{\rm m}^{-2}\,{\rm sr}^{-1}$. The green thick lines show the default \cite{Helgason:2012a} reconstruction of the contributions from the remaining known galaxies with the range marked and bounded by the green dotted lines. The reconstruction has 1-Halo (1H) and 2H (clustering) components \citep{Cooray:2004,Zheng:2005,Helgason:2012a}. The figure shows that 1) large scales ($\gtrsim 200\arcsec$) cannot be accounted for by the known remaining galaxies indicating contributions from the new sources, and 2) the 1H component from the known galaxies appears always non-negligible, and at times dominant, even close to the beam scales, which has other implications outside the scope of this discussion. In discussions below we are mainly interested in the amplitude of the CIB fluctuation excess and thus do not correct in detail for the mask effects, as e.g., discussed and done in \cite{Kashlinsky:2005a} and \cite{Kashlinsky:2012}, which is important when making precision fits of theoretical templates to the data. This will be presented, for the various specific map masks, later when fitting specific templates.

\begin{figure}[ht!]
 \includegraphics[width=6in]{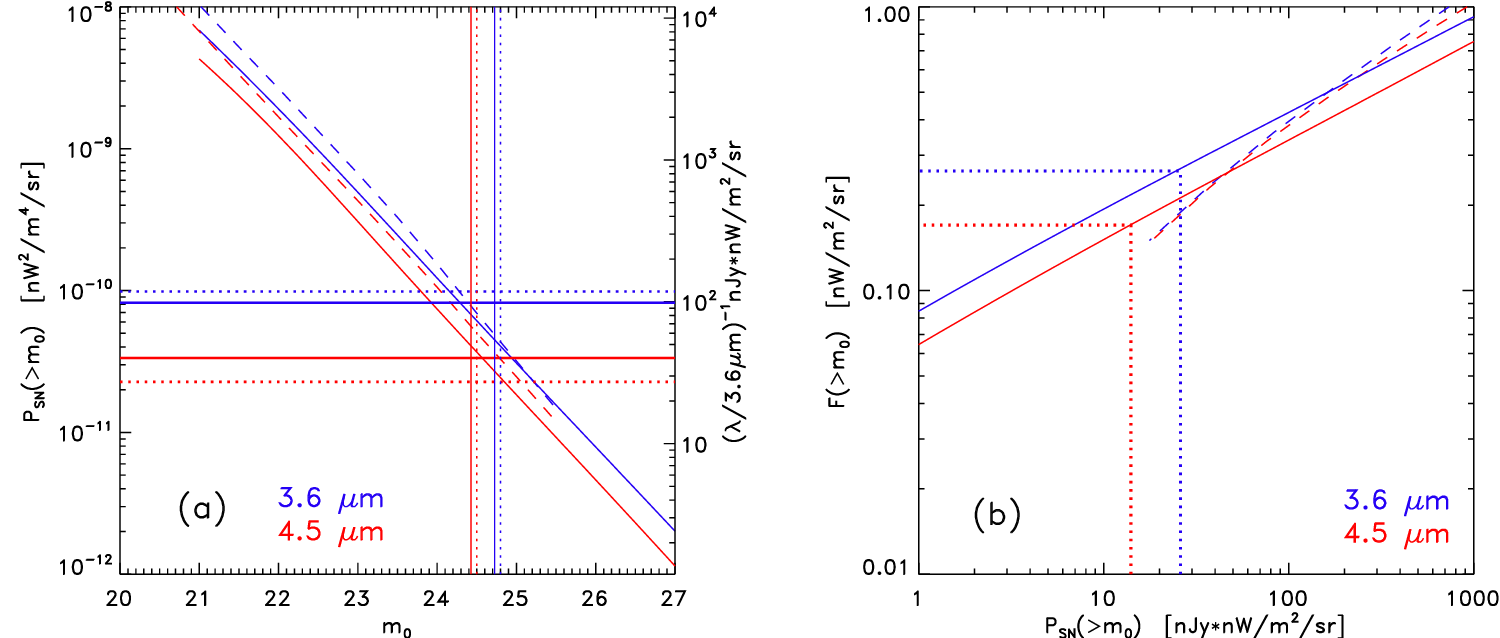}
\caption{(a) $P_{\rm SN}$ from known galaxies with AB magnitude $>m_0$ from JWST counts \citep[solid lines,][]{Windhorst:2023} and Spitzer \citep[dashed lines,][]{Ashby:2013,Ashby:2015}. The shot noise levels remaining in the fields are shown with horizontal lines: blue, red at 3.6, 4.5 \mic.   
  The adopted magnitude of source clipping is shown with vertical lines: $m_{\rm AB}=24.8, 24.5$ at 3.6, 4.5 \mic,  which are slightly shifted for clearer display. Vertical and horizontal lines correspond to NEP (dotted) and CDFS (solid).
  (b) Flux remaining from known galaxies of $m>m_0$ vs the shot-noise contributed by them. Vertical blue/red lines show $P_{\rm SN}$ at 3.6/4.5 \mic\ from Table 1 of \cite{Kashlinsky:2007a} marking also the $F(>m_0)$ due to  these galaxies.}
\label{fig:fig2b}
\end{figure}
The expected remaining flux and shot-noise power are related to the source counts: $F(>\!\!m_0)=\int_{m_0}^\infty S(m) (dN/dm) dm$ and $P_{\rm SN}(>\!\!m_0)=\int_{m_0}^\infty S^2(m) (dN/dm) dm$, where $S(m)=3631 \cdot 10^{-0.4m}$Jy. The two are interrelated via $P_{\rm SN} =\bar{S} F$ with $\bar{S}$ being the suitably averaged individual flux of the sources. Figure \ref{fig:fig2b}a reconstructs the shot noise, $P_{\rm SN}(>\!\!m_0)$, over the magnitude range observed with Spitzer \citep[][]{Ashby:2013,Ashby:2015} and JWST \citep[][]{Windhorst:2023} shown with blue (3.6 \mic) and red (4.5 \mic) colors. The horizontal lines show the effective shot-noise power in the two regions. The vertical lines show the magnitudes corresponding to the 3$\sigma$ flux level of the final maps; they are similar but slightly shifted for clarity. The level of the equivalent shot-noise power fitted empirically here carries systematic uncertainty and is subdominant to the 1H component, so it represents the upper limit on the true shot-noise power from the sources remaining in the maps.

\begin{figure}[ht!]
\includegraphics[width=5.25in]{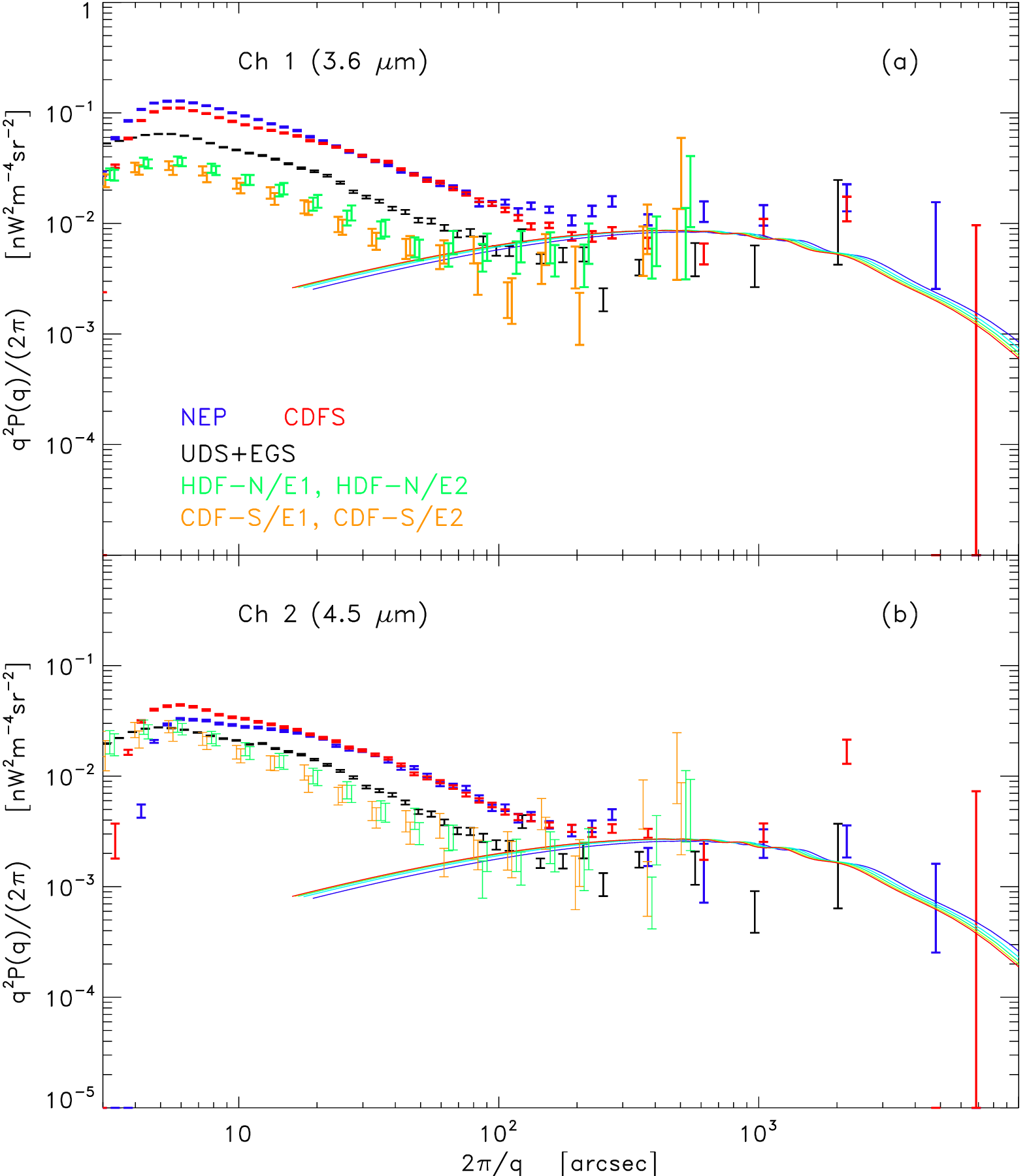}
 \caption{Source-subtracted CIB anisotropies from the analysis here (blue and red), from Fig. 9 of \cite{Kashlinsky:2012} (black) and from Fig. 1 of \cite{Kashlinsky:2007a} (green and brown). The powers here and at \cite{Kashlinsky:2012} are evaluated at the same $q$ grid, but black symbols are slightly shifted for clarity. The brown/green data for the $q$-grid from \cite{Kashlinsky:2007a,Kashlinsky:2007b} was slightly shifted for clearer display with the leftmost green corresponding to the original $q$.  The lines correspond to single-$z$ templates of the CIB at $z=10,15,20,25,30$.}
\label{fig:fig3}
\end{figure}
Figure~\ref{fig:fig3} compares these measurements with each other and with the prior probes by \cite{Kashlinsky:2007a,Kashlinsky:2007,Kashlinsky:2012} at lower shot noise levels but smaller scales. Fig.\ \ref{fig:fig3} also shows the good agreement between the fields  at various Galactic latitudes indicating no significant cirrus pollution in CIB. The Limber equation \citep[e.g., see Appendix in][]{Fernandez:2010} gives CIB fluctuations from sources populating the redshift cone:
\begin{equation}
q^2P(q) = \int \frac{(\partial F/\partial z)^2}{cH^{-1}(z)} \left[(q/d_A)^2{\cal P}_{\rm 3D}(q/d_A)\right] dz
\label{eq:limber}
\end{equation}
where $d_A(z)=c\int_0^z H^{-1}(z) dz$ and ${\cal P}_{\rm 3D}$ is the underlying 3-D power spectrum of the emitting sources. When CIB emissions span a narrow range of $z$ one gets $q^2P(q)\propto [x^2{\cal P}_{\rm 3D}(x)]|_{x=qd_A^{-1}}$. 
As an example the spectrum from such sources at fixed $10\leq z\leq 30$ values is shown in Fig. \ref{fig:fig3} where ${\cal P}_{\rm 3D}$ was adopted from CMBFAST \citep{Seljak:1996} with standard cosmological parameters. 
The modeled ${\cal P}_{\rm 3D}$ are generally consistent with the data, which are currently
insufficient to distinguish between the various $z$ contributions.

General considerations resulting from the amplitude of the CIB fluctuations by sources below the probed shot-noise level were discussed by \cite{Kashlinsky:2007b,Kashlinsky:2018}: given the amplitude of the CIB fluctuations ($\delta F\sim 0.05-0.1$ nW/m$^2$/sr on arcminute to sub-degree scales), it is very hard to produce the observed fluctuation from the remaining known sources that can contribute $F(>\!\!m_0) < 0.15-0.2$ nW/m$^2$/sr in the total flux as shown in Fig. \ref{fig:fig2b}b. The net flux required to be produced by the CIB sources is $F_{2-5\mic}\sim 1$ nW/m$^2$/sr and it has to be produced by very faint sources with individual fluxes of typically $\bar{S}< P_{\rm SN}/F_{2-5\mic}\sim$0.15--0.2 nJy which are below the confusion noise of Spitzer/IRAC. The earlier measurements reaching lower shot-noise levels by \cite{Kashlinsky:2007a,Kashlinsky:2007,Kashlinsky:2012}, which are deeper but cover a smaller range of angular scales,  are shown for comparison in Fig. \ref{fig:fig3}. The fact that there is no obvious decrease of the clustering component from the new sources implied by the CIB fluctuations with the decreasing shot-noise power likewise indicates that the new sources lie at magnitudes not yet accessible here individually.

We now turn to discussing the particular sources that could, or were proposed to, contribute to the measured CIB fluctuation signal at 3.6 and 4.5 \mic. 
\begin{enumerate}
\item Remaining/unresolved known galaxy populations below the shot-noise reached here were shown already in the original study to produce CIB fluctuation well below the measured levels \citep[][]{Kashlinsky:2005a}. The detailed machinery developed by \cite{Helgason:2012a} from a compilation of galaxy luminosity functions at various wavelengths and redshifts allowed to robustly fill in the redshift cone to $z\sim 6$ with the standard $\Lambda$CDM density fluctuation spectrum and evaluate the CIB anisotropies from known sources below the imposed cutoffs in either shot-noise, magnitude or both. 
The reconstruction provides an excellent fit to the galaxy counts measured to much fainter limits after its development as can be seen by comparing Fig. 5 of \cite{Helgason:2012a} and Fig. 6 of \cite{Kashlinsky:2024}. The galaxy clustering includes the 1H and 2H components  \citep{Cooray:2002,Zheng:2005} which at the appropriate shot-noise and achieved magnitude limits lead to CIB anisotropies well below the measured level at $2\pi/q\gtrsim 5\arcmin$. Thus the remaining known galaxies cannot explain the CIB fluctuation signal in Figs. \ref{fig:fig4} and \ref{fig:fig3} now measured to $\gtrsim 1\deg$.
\item Population III and early emissions were proposed to be detectable via their contributions to the near-IR CIB anisotropies \citep{Cooray:2004,Kashlinsky:2004} and this was originally the motivation for the search of \cite{Kashlinsky:2005a}. However, while the required fraction appears small, as discussed in \cite{Helgason:2016} and \cite{Kashlinsky:2018}, with the small-scale power of the {\it standard} $\Lambda$CDM model this component would require a significantly high efficiency of star formation inside the collapsed high-$z$ halos. Indeed, populations that have converted fraction $f$ of the available cosmic baryons into radiation with the H-burning efficiency $\epsilon=\epsilon_{\rm H} = 7\times 10^{-3}$ (reached in massive, fully convective stars \citep{Schaerer:2002} such as Pop III) redshifted into the observer near-IR frame from the typical redshift $z$ would produce CIB flux of $F=f\epsilon \frac{c}{4\pi}\rho_{\rm baryon}c^2(1+z)^{-1}\sim630 f (\epsilon/\epsilon_{\rm H})[(1+z)/10]^{-1}$ nW/m$^2$/sr \citep{Kashlinsky:2005b,Kashlinsky:2015a}. So to produce $F_{\rm CIB}\sim 1$ nW/m$^2$/sr by $z\sim 10$ would require a fraction $f\sim 1.6\times 10^{-3}[(1+z)/10]$ of all the baryons to have gone through stars by that $z$. Additional emissions from supernovae expected to come abundantly from such massive stars would decrease the required here $f$ further \citep{Kashlinsky:2015a}. Given that Pop III are expected to be very massive, the CXB-CIB cross power would then be presumably generated from the sufficient fraction of those stars ($\sim 10-20\%$) leaving black hole remnants producing coexisting CXB emissions as discussed in e.g. \cite{Helgason:2016}; this is one of the differences between Pop III vs present-day stellar populations viz-a-viz the observational CIB data. Nevertheless the implied (low) value of $f$ may be still hard to achieve from the early stars era in the standard $\Lambda$CDM model where the small scale fluctuation power from inflationary density field  decreases as ${\cal P}_{\rm 3D}\propto k^{-3}$ leading to relatively low abundance of collapsed halos suitable for star formation at $z>10$  \citep{Helgason:2016}; the spectral index ($-3$) in the small-scale ${\cal P}_{\rm 3D}$ reflects the 3-dimensionality of space in the Universe and is robust. However, the required $f$ can be produced more naturally if an extra small scale power is added to the inflation-produced ${\cal P}_{\rm 3D}$ such as if LIGO-type PBHs make up the dark matter discussed below. 

\item IHL: It was suggested that the intrahalo light (IHL) from normal stars detached from parent galaxies at low and intermediate redshifts can produce the CIB anisotropies detected in Spitzer studies \citep{Cooray:2012,Zemcov:2014}. The stars here would be at lower $z$, but at the same time the effective efficiency $\epsilon$ would be over an order of magnitude smaller than $\epsilon_{\rm H}$ as only a small core there would be burning hydrogen. As soon as this proposal appeared \citep{Cooray:2012}, the CXB-CIB cross-power was detected \citep{Cappelluti:2013}. No explanation has been proposed to date for that observation and the levels of the cross-power and the consequent CXB-CIB coherence levels appear to rule out the IHL models as produced by present-day (normal) stellar populations.  Additional issues with the data analysis there were pointed out to be related to the large areas of the studied maps ($>70\%$) removed prior to FT, and it was proposed to compute the maps' correlation function, which is immune to masking, to show consistency with the claimed CIB power from Fourier analysis \citep{Kashlinsky:2015a}; no such calculation has appeared as of this writing.  
\item DCBH: Alternatively if the emitted radiation is produced by black hole accretion processes reaching much greater $\epsilon$, the CIB flux per baryon would be proportionately higher. Following the discovery of the CIB-CXB cross-power by \cite{Cappelluti:2013} it was suggested by \cite{Yue:2013} that the role of the direct collapse black holes (DCBHs) at $z\sim 10$ may be important in producing the observed signals. Such DCBHs, the accretion onto which can potentially reach $\epsilon=0.4$, as in maximally rotating Kerr black holes, would form by direct collapse of early halos and generate the required emissions as further developed in \cite{Yue:2014,Yue:2016a}. With the much higher efficiency, 
the fraction of baryons needed to produce the required CIB with this mechanism would be reduced to $f\sim 10^{-4}(\epsilon/0.1)^{-1}[(1+z)/10]$. In this model the CIB anisotropies are proposed to come from a population of highly obscured DCBHs formed in metal-free halos with virial temperature above $10^4$ K at $z \gtrsim 12$ \citep{Yue:2013}. DCBHs may well be a consequence of the Pop III era and evolution, so this interesting model may in a way be tied to the general high-$z$ era discussions with Pop III also contributing to the overall $f$ (now reduced thanks to DCBH contributions) and CIB levels.
\item PBH-DM: Following the first LIGO gravitational wave event GW150914 \citep{Abbott:2016b} it was proposed that LIGO type primordial black holes (PBHs) constitute the bulk of the dark matter (DM) and hence naturally reproduce the observed CIB anisotropies while also accounting for their high coherence with the unresolved CXB \citep{Kashlinsky:2016}. See \cite{Carr:2024} and references therein for review on the PBH-DM connection. The CIB component would then reflect the additional white-noise \cite[from granulation,][]{Meszaros:1974} and model-independent component in the matter power spectrum. At mass scales relevant for early forming halos, that component dominates the ${\cal P}_{\rm 3D}\propto k^{-3}$ term from inflation, leading to significantly increased abundance of star forming collapsed halos at $z>10$, producing the observed CIB excess. The required $\sim 0.1\%$ of the baryons to account for it would easily come from the significantly more collapsed halos at $z>10-20$ due to the granulation power component. The CXB-CIB coherence provides further evidence for black holes at early times (such as Pop III remnants, DCBH, and/or - but not necessariyly - PBH) with emissions by the accreting PBHs modeled in this context by \cite{Hasinger:2020} and \cite{Cappelluti:2022}. This proposal would also be consistent with efficient gas collapse and subsequent star formation at high-$z$ \citep{Kashlinsky:2021,Atrio-Barandela:2022}. This, to date, appears to be the most successful explanation of the CIB anisotropies discussed here. The LIGO-Virgo-KAGRA O5 run at significantly increased sensitivity, reaching $\sim$300 Mpc, is planned to start ca. 2027 (\url{https://www.ligo.caltech.edu/page/observing-plans}), bringing new data regarding the putative PBH-DM connection.
\end{enumerate}

\section{Future prospects}

This dataset and its results are important for future CIB studies advancing understanding of the origin of the source-subtracted CIB anisotropies. In particular:

The CIB maps here offer a new important opportunity to probe the CXB-CIB cross-power to large(r) scales, as can be done with spatially overlapping ROSAT, Chandra, XMM-Newton and eROSITA data in the soft X-ray bands near 1--2 keV. In addition to probing significantly larger angular scales than before, such measurement will be useful for further increasing the overall S/N of the cross-power, potentially identifying spatial spectrum features that could be important for probing the epochs of the sources.

Appropriate space-based new missions surveying substantial sky areas at adjacent wavelengths would bring new advances here. To further probe reliably emissions from faint populations at high $z$ from suitably prepared maps one would want 1) to cover larger areas, 2) to achieve sufficient depth, $m_{\rm AB}\gtrsim 24-25$ \citep{Kashlinsky:2004}, 3) subarcsecond angular resolution (to identify/remove galaxies while leaving enough sky for robust FT), and 4) data at shorter IR bands along with visible data which would give direct indication of the Ly-break in the energy spectrum.  This can and will be achieved with the Euclid's Wide Survey covering 15,000 deg$^2$ which will also have $\sim 100$ deg$^2$ of Deep Survey going two magnitudes fainter \citep{Mellier:2024}.  The latter is important to potentially probe the shot-noise component originating from the new populations to gain insights into their individual fluxes \citep{Kashlinsky:2018}.  Additional methodology for this has been proposed by \cite{Kashlinsky:2015} for the Lyman tomography to isolate emissions by their epochs from several filters of adjacent wavelengths as in Euclid's Near-Infrared Spectrometer and Photometer (NISP), also applicable to Roman.
The Roman mission survey designs have not been finalized as of this writing. It is anticipated that these will cover several thousand square degrees to a depth one to two magnitudes fainter than the Euclid wide survey, and several tens of square degrees to still greater depths. The wide area survey will likely be split, with part observed in two near-IR bands and part observed in 3 near-IR bands. The deep survey regions will be observed in more bands, likely four in the near-IR to as many as eight that span the full 0.48-2.3 $\mu$m wavelength coverage available to the Roman wide-field instrument. The greater angular resolution and depth of Roman will provide excellent identification and removal of stars and galaxies from the images. 
Both Euclid and Roman have visible wavelengths covered to probe the Lyman break of the CIB anisotropies.

 \begin{acknowledgments}
 Work by A. K. and R. G. A. was supported by NASA under award number 80GSFC24M0006. The authors acknowledge support from NASA award 80NSSC22K0621 
 ``Precision measurement of source-subtracted cosmic infrared background from new Spitzer data". 
 This work is based on observations made with the Spitzer Space Telescope, which was operated by the Jet Propulsion Laboratory, California Institute of Technology under a contract with NASA. We thank Kari Helgason for the software for the LIBRAE project (\url{https://euclid.caltech.edu/page/kashlinsky-team}) to reproduce CIB anisotropies from known galaxy populations. 
 \end{acknowledgments}

\section*{APPENDIX}
{\bf Image processing}. 
\begin{figure*}[ht!] 
   \centering
  \includegraphics[height=1.9in] {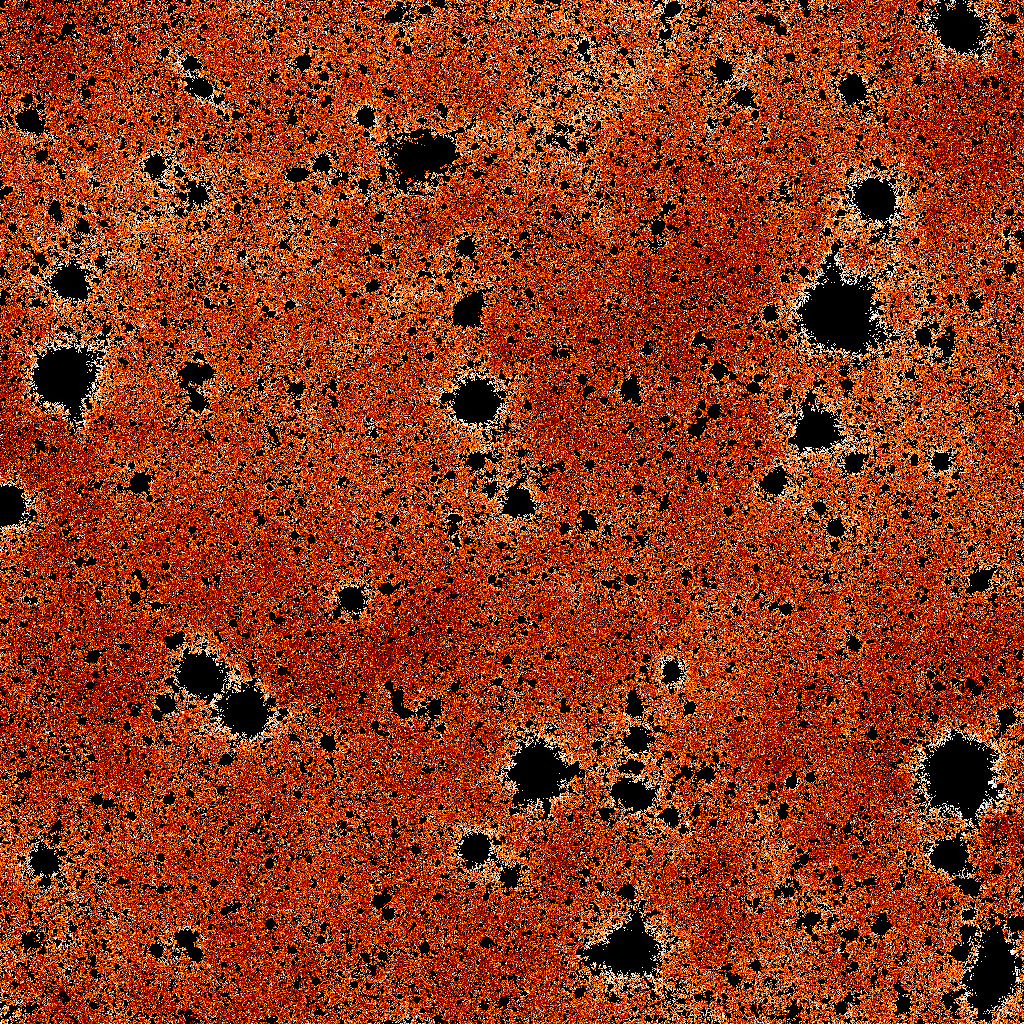} ~
   \includegraphics[height=1.32in]{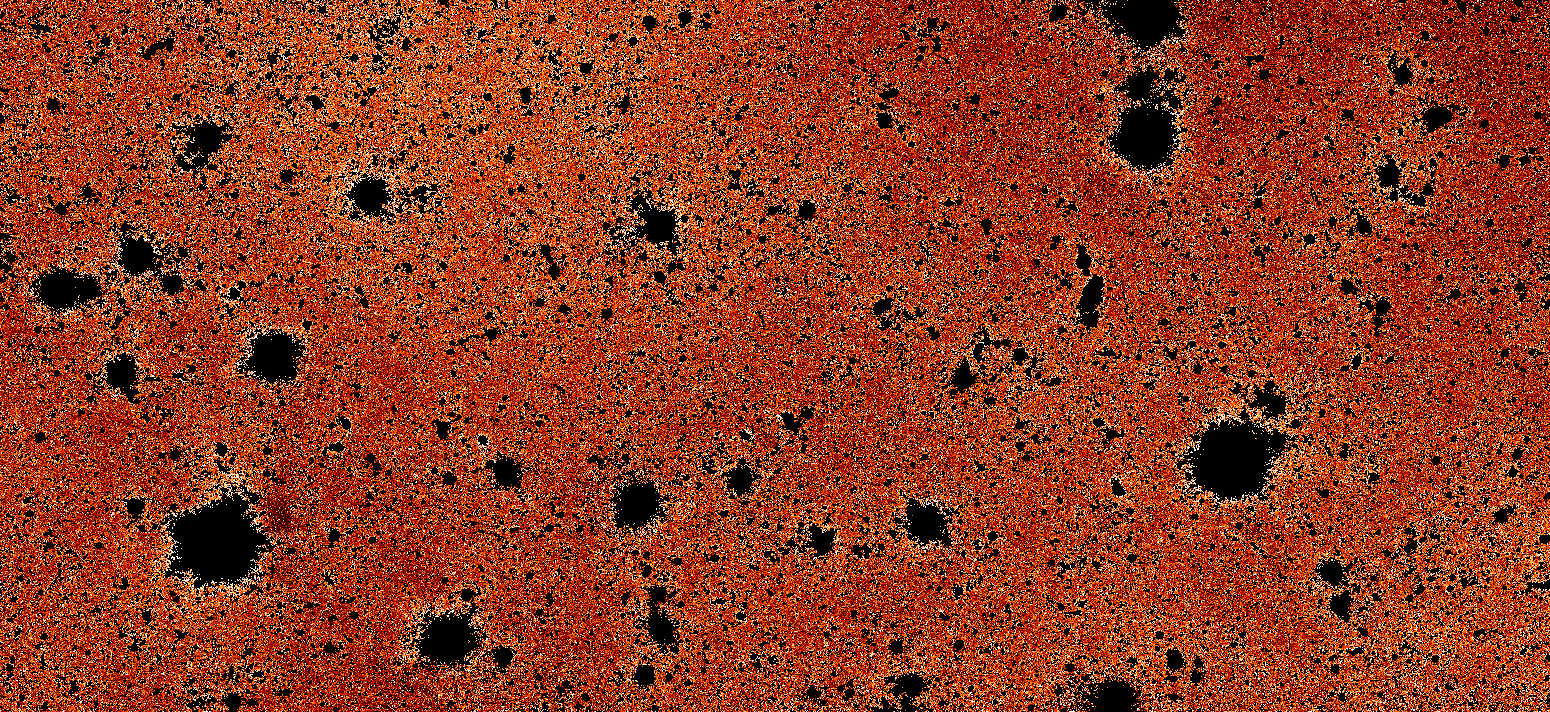} ~~~
   \includegraphics[height=1.9in]{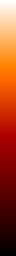}\\
   ~~\\
   \includegraphics[height=1.9in] {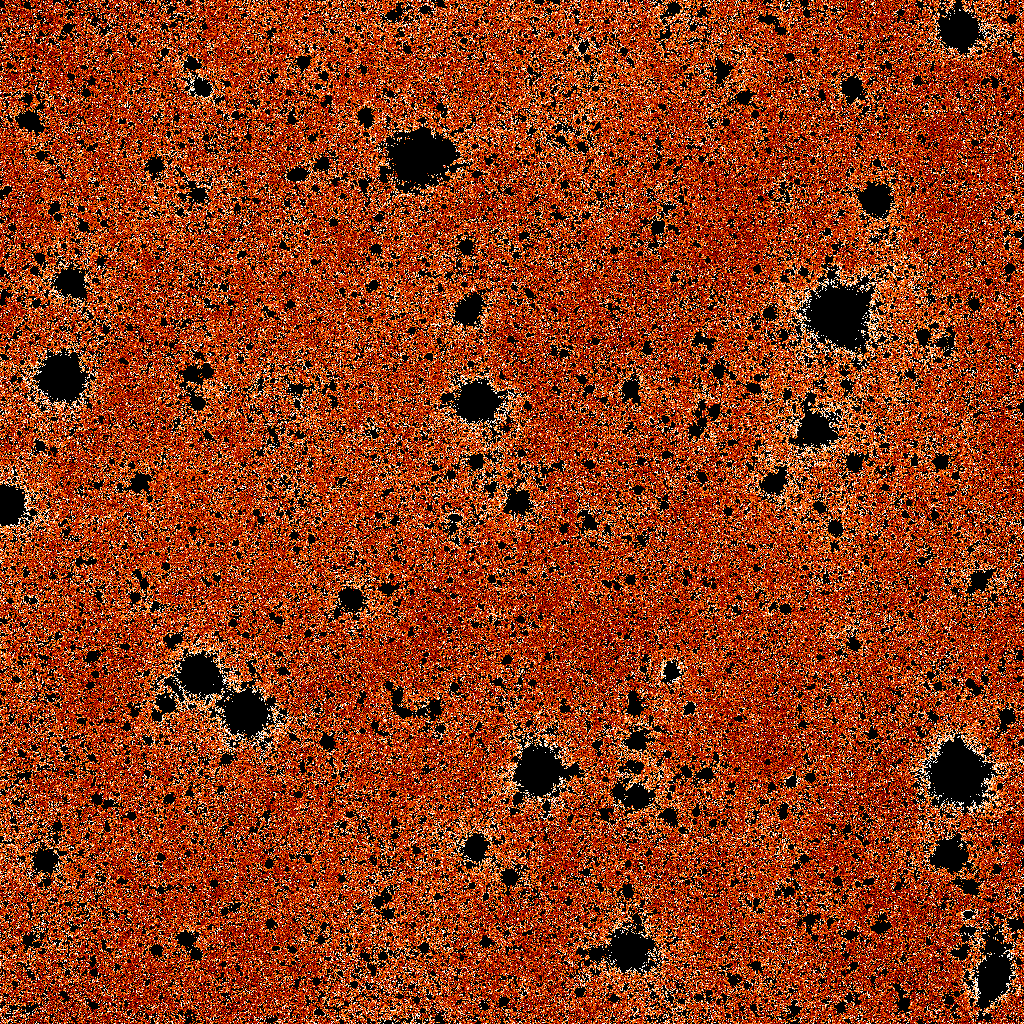} ~
     \includegraphics[height=1.32in]{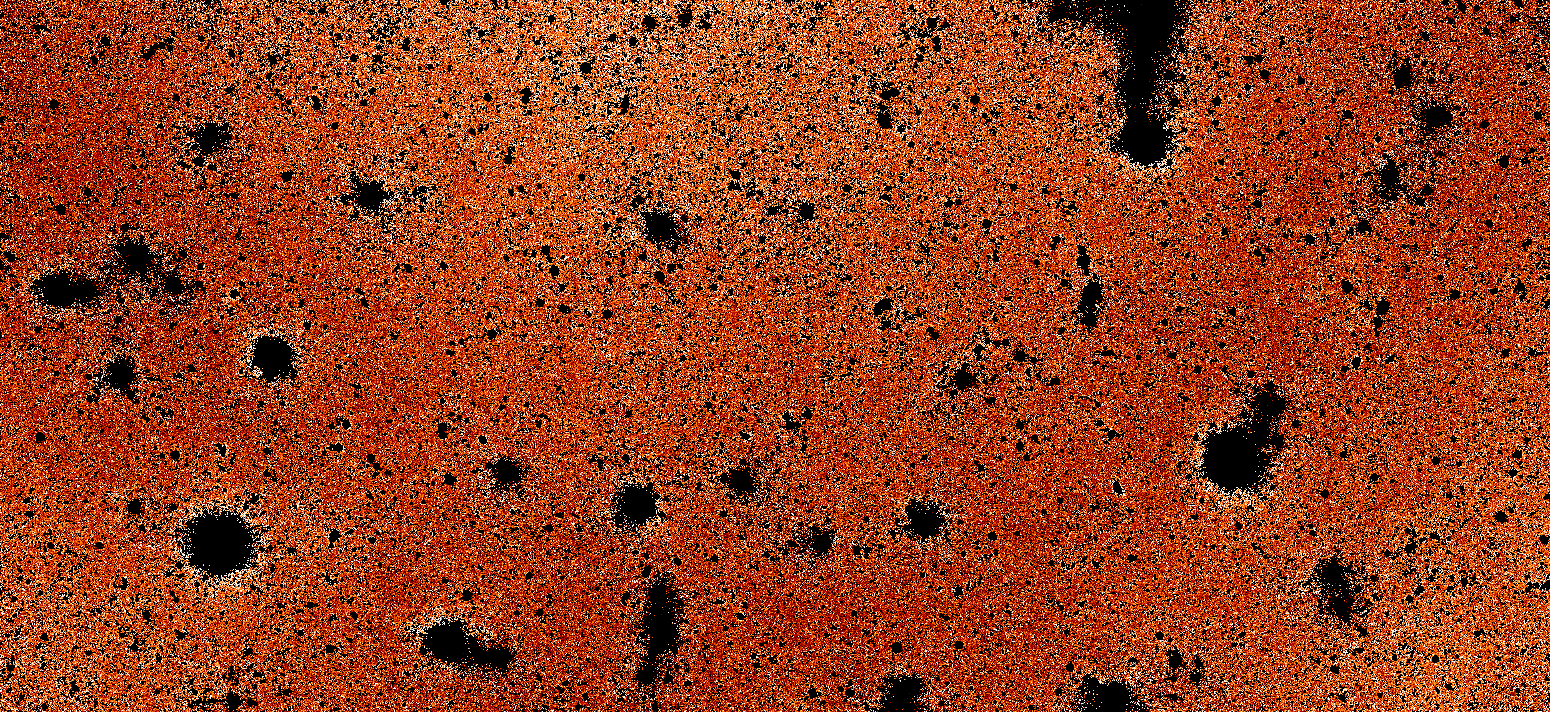} ~~~
   \includegraphics[height=1.9in]{bar}\\
   \vspace{0.7cm}
\includegraphics[width=6in]{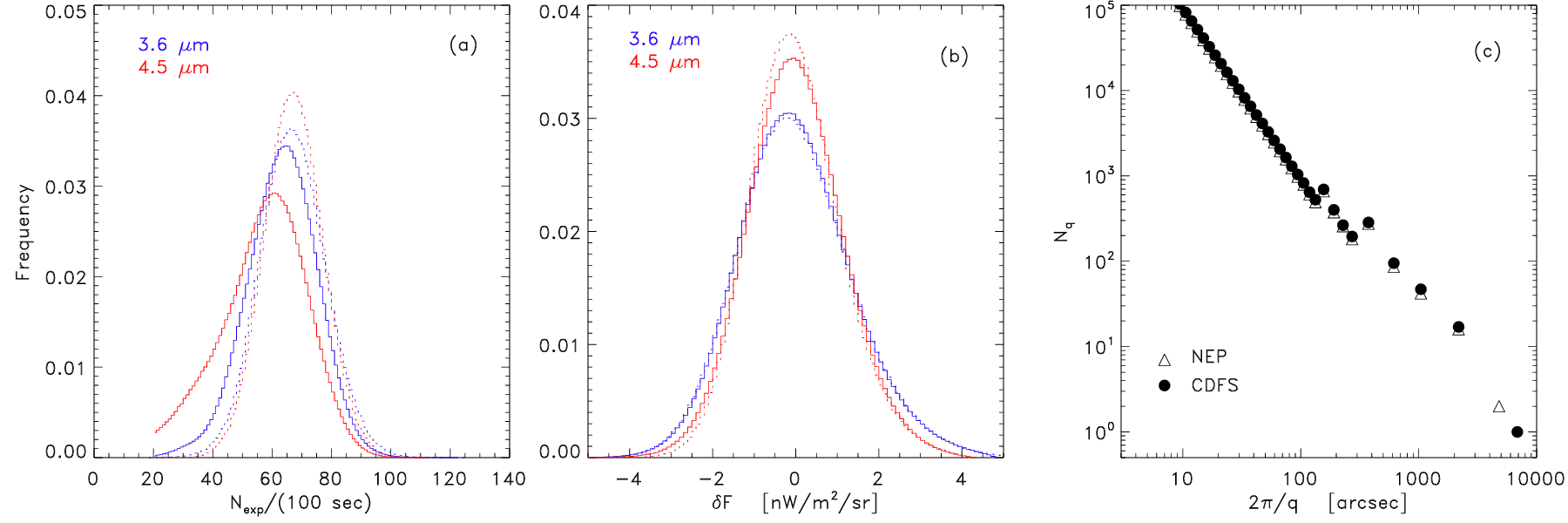}
\caption{NEP (left) and CDFS (right) fields after masking and source subtraction as described.
The 3.6 $\mu$m images (top row) are scaled [$-4.2$, 2.5] nW~m$^{-2}$~sr$^{-1}$.
The 4.5 $\mu$m images (bottom row) are scaled [$-3.3$, 2.0] nW~m$^{-2}$~sr$^{-1}$.
For display purposes these images have been resampled to $4\farcs8$ pixel sizes.
The truncation of an extended bright region by the top edge of the CDFS 4.5 $\mu$m image 
causes the unusually high power at $\sim 2000''$
in Figs.\ \ref{fig:fig2} and \ref{fig:fig4}.
BOTTOM: Left: Exposure times for the selected for CIB analysis regions. Red/blue are 3.6/4.5 \mic; solid/dotted are CDFS/NEP fields. Middle: flux histograms at 3.6 (left) and 4.5 (right) \mic; solid/dotted are CDFS/NEP fields. Right: The number of independent Fourier elements, $N_q$, to compute power at each $q$ with the division of the Fourier space here per \cite{Kashlinsky:2012}. Blue/red correspond to NEP/CDFS fields.
   \label{fig:fields}}
\end{figure*}
The self-calibration method is now standard for such studies as discussed in great detail in \cite{Arendt:2010}. Here we use the same data model
$D^i = S^\alpha + F^p + F^q$ where $D^i$ is a single datum from one pixel of one exposure,
$S^\alpha$ is the sky intensity at the location $\alpha$ 
observed by that pixel, $F^p$ is the detector offset for pixel $p$,
and $F^q$ is an offset for each of the four IRAC detector outputs that is allowed 
to vary from frame to frame. A difference from the processing in 
\cite{Arendt:2010} is that here the $F^p$ term was allowed to vary 
as a function of the delay time since the previous exposure in addition 
to the previous variation for each AOR. This allows the 
self-calibration to compensate for 
IRAC's ``first frame effect'' \citep{Fazio:2004}.
For the CDFS field, the self-calibration of the data omitted exposures
containing the brightest stars in the field because abnormally large
offsets in these exposures are not well tracked in the data model, 
and an extended region around these very bright sources is to be masked regardless.
The self-calibrated data were mapped into sky maps with a $1.2''$ pixel scale. 
Figure \ref{fig:fields} shows images of the regions selected for Fourier analysis. The 2-D FTs for the CDFS show a stripe through the vertical axis that is not 
present for the NEP. The stripe is generated because there is an extended 
bright region in the CDFS maps that is sharply truncated by the edge of 
the selected field (top edge, left of center in Fig. \ref{fig:fields}). The 
excess power in this stripe contributes to the elevated point in the CDFS
fluctuation spectrum at $\sim2000''$ (especially at 4.5 $\mu$m).

{\bf Noise evaluation}.
\begin{figure}[ht!]
\includegraphics[width=5.8in]{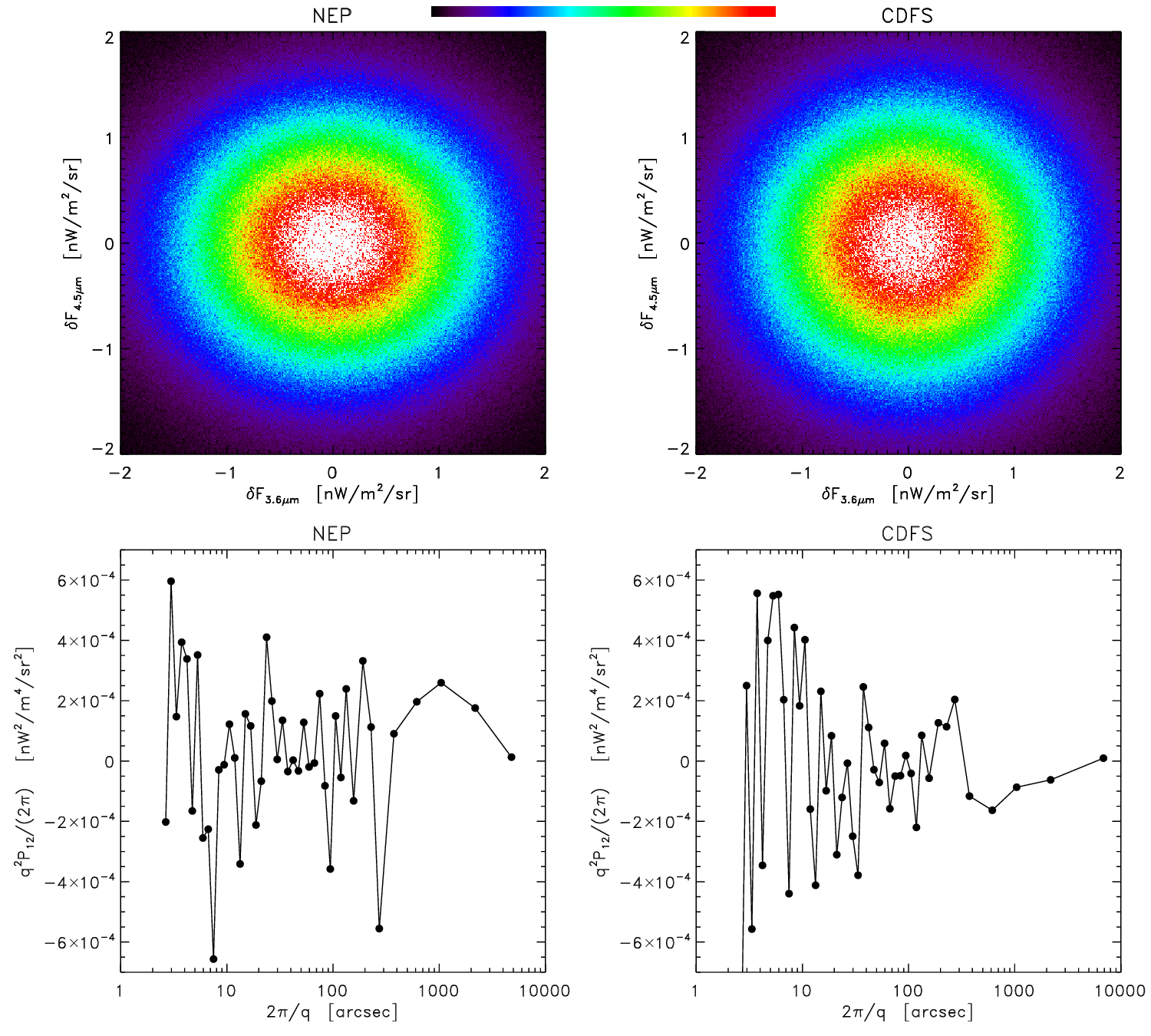}
 \caption{Noise properties for the $A-B$ maps: top shows the scatter plot between the two wavelengths illustrating a completely random, uncorrelated components between the two wavelengths for each of the fields. Bottom shows the cross-powers between the two wavelengths in each of the fields demonstrating randomness between the two wavelengths.}
\label{fig:fig5}
\end{figure}
In principle, construction of the $A-B$ difference map 
removes all astronomical sources and leaves a map showing the characteristics
of the random noise in the data. However, because these observations were
performed using 2 repeats at each location, two subsets generated by assigning 
individual exposures alternately to the $A$ and $B$ subsets may tend to be 
dominated by either first or second exposures with different delay times 
and offsets. Occasionally, an exposure is missing or excluded, which 
switches the subsequent association between the first and 
second exposures and the two subsets. This adds some spatially correlated systematic 
differences in the $A$ and $B$ maps that are exaggerated in the difference map. 
Thus to obtain a more uniform noise map, we randomly (rather than in 
temporal order) assigned exposures to $A$ and $B$ subsets. These $A-B$ noise maps 
have a weaker imprint of the observing sequence, but have more variation 
in the depth of coverage between the separate $A$ and $B$ mosaics. 
Figure \ref{fig:fig5} demonstrates the independence of the evaluated noise 
between the two wavelengths consistent with the expectation from noise.

{\bf Foreground contributions: zodiacal light and cirrus}.
Zodiacal light may contribute to the white noise component, and the amplitude of this contribution can be deduced from \cite{Kelsall:1998} and \cite{Arendt:2016}. The ranges of their contributions are plotted with blue vertical dashes in Fig. \ref{fig:fig2}. This contribution appears small on scales exceeding the beam here. 

\begin{figure}[ht!]
\includegraphics[width=6.5in]{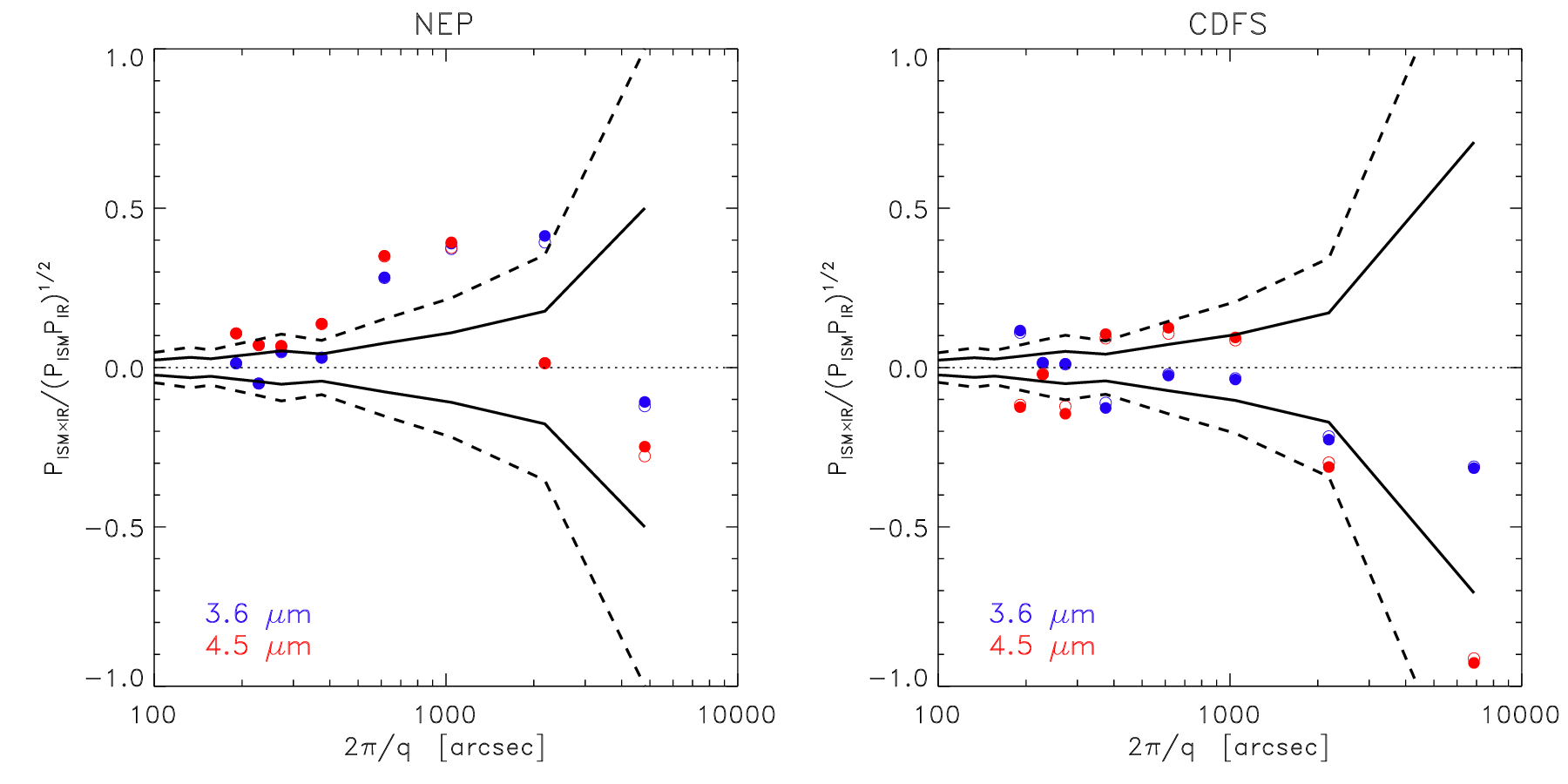}
 \caption{The dimensionless cross-power between the IRAS 100 \mic\ cirrus data and the 
 Spitzer IR maps for each of the two fields under study. Filled and open circles represent 
 use of either the ISSA or IRIS reductions
 of the IRAS survey. Solid and dashed lines indicate the 
 ranges of $1/N_q^{0.5}$ and $2/N_q^{0.5}$ that would be expected from uncorrelated data to reflect 1 and 2 $\sigma$ at small ${\cal R}$. [Note that this does not strictly reflect the appropriate confidence levels and as ${\cal R}$ approaches unity the latter must be evaluated from the Fisher transformation as discussed in \cite{Kashlinsky:2018}]. The amplitude of the cross-power demonstrates the negligible contribution of cirrus to the power displayed in Fig. \ref{fig:fig4}.}
\label{fig:fig_cirrus}
\end{figure}

To evaluate possible contributions from dust in the ISM (cirrus), we 
use the 100~\mic\ IRAS data \citep{Neugebauer:1984} and compute the cross-power, $P_{{\rm 
ISM}\times{\rm IR}}$, between that map and the IR data here. Fig. \ref{fig:fig_cirrus} 
shows ${\cal R}\equiv P_{{\rm ISM}\times{\rm IR}}/\sqrt{P_{\rm ISM}P_{\rm IR}}$ at 
$>180\arcsec$ scales, limited by the $90''$ pixel of the original 
100~\mic\ map; the coherence is ${\cal C}={\cal R}^2$ \citep[see Fig. 8][]{Kashlinsky:2012}. Assuming spatially-constant color of Galactic cirrus from 100 \mic\ to the IRAC bands
[e.g. such that $P_{\rm ISM}(100\ \mu{\rm m}) \propto P_{\rm cirrus}(3.6,4.5\ \mu{\rm m})$],
and the absence of Galactic absorption at the Galactic latitudes leads to 
$P_{\rm cirrus}/P_{\rm IR}={\cal C} = 0.16({\cal R}/0.4)^2$, showing negligible cirrus contributions to the IR power per Fig. \ref{fig:fig_cirrus}. For the CDFS, the values of ${\cal R}$ are consistent with 
the results that would be expected from random uncorrelated data, although the 100 $\mu$m signal 
is extremely faint here. In the NEP field, the ISM emission is brighter and clearly 
detected in the IRAS images. There is indication of a possible weak correlation 
with the Spitzer data here, but the amplitude and significance are low, 
suggesting that the CIB fluctuation measurements are not affected
by ISM emission.

\clearpage


\end{document}